\documentclass[sn-mathphys-num]{sn-jnl}

\usepackage{graphicx}
\usepackage{multirow}%
\usepackage{amsmath,amssymb,amsfonts}%
\usepackage{amsthm}%
\usepackage{mathrsfs}%
\usepackage[title]{appendix}%
\usepackage{xcolor}%
\usepackage{textcomp}%
\usepackage{manyfoot}%
\usepackage{booktabs}%
\usepackage{algorithm}%
\usepackage{algorithmicx}%
\usepackage{algpseudocode}%
\usepackage{listings}%
\usepackage{float}%
\usepackage{placeins}%
\usepackage{tabularx}
\usepackage{array}        %
\usepackage{caption}      %

\theoremstyle{thmstyleone}%
\theoremstyle{thmstyletwo}%

\theoremstyle{thmstylethree}%

\usepackage{pgfplots}
\usepackage{pgfplotstable}
\usepgfplotslibrary{statistics}
\pgfplotsset{
    compat = 1.18,
    only if/.style args={entry of #1 is #2}{
        /pgfplots/boxplot/data filter/.code={
            \edef\tempa{\thisrow{#1}}
            \edef\tempb{#2}
            \ifx\tempa\tempb
            \else
                
            \fi
        }
    }
}

\begin{document}

\title[Revealing Quantum Information Encoded in Classical Images]{Revealing Quantum Information Encoded in Classical Images}


\author*[1]{\fnm{Otmane} \sur{Ainelkitane}}\email{Otmane.Ainelkitane@canberra.edu.au}

\author[1]{\fnm{Brian } \sur{Recktenwall-Calvet}}\email{riverscripttechservices@gmail.com}

\author[1]{\fnm{Aasma } \sur{Iqbal}}\email{aasmaiqbal.iqbal@canberra.edu.au}

\author*[1]{\fnm{Carlos C.} \sur{N. Kuhn}}\email{carlos.noschangkuhn@canberra.edu.au}



\affil[1]{\orgdiv{OpenSI - Faculty of Science and Technology }, \orgname{University of Canberra}, \orgaddress{\street{11 Kirinary Street}, \city{Bruce}, \postcode{2617}, \state{ACT}, \country{Australia}}}



\abstract{In this study, we investigate a simple quantum pre-processing filter kernel designed with only two CNOT gates for image feature extraction. We examine the impact of these filters when combined with a classical neural network for image classification tasks.
Our main hypothesis is that this circuit can extract pixel correlation information that classical filters cannot. This approach is akin to a convolutional neural network, but with quantum layers replacing convolutional layers to extract spatial pixel entanglement. We found that a small circuit with just two CNOT gates can be engineered in three different spatial symmetries, each affecting classification differently. While the filter improves classification when combined with a simple, narrow network, it does not surpass complex classical methods. However, the filter demonstrates potential to enhance classification performance in more sophisticated architectures. Despite this, our empirical results show no clear correlation between the observed improvements and the level of entanglement in the quantum circuit, as measured by Von Neumann Entropy. The underlying cause of this improvement remains unclear and warrants further investigation.
}




\maketitle


\section{Introduction}

The rapid advancements in both machine learning (ML) and quantum computing have led to the emergence of quantum machine learning (QML), a promising interdisciplinary field that leverages quantum mechanics to enhance computational capabilities \cite{BiamonteReview2017}. Among the many applications of ML, image classification has seen remarkable progress in recent decades, becoming increasingly crucial across diverse sectors, including medical diagnosis and autonomous driving. Convolutional Neural Networks (CNNs) \cite{lecun1998gradient} have played a central role in this success, excelling at hierarchically extracting spatial features from visual data.

Numerous researchers have developed various concepts for QML. Early work explored quantum algorithms for support vector machines (SVMs), demonstrating potential speedups over their classical counterparts \cite{LloydSVM2013}. Initial theoretical explorations also investigated how fundamental neural‐network components, such as artificial neurons (i.e., the Rosenblatt perceptron \cite{RosenblattPerceptron}), could be realised on quantum hardware \cite{Kak1995}. More recently, Tacchino \emph{et~al.}\ revisited the quantum perceptron using parameterised quantum circuits (PQCs) with hardware-efficient single-qubit gates that can be stacked into trainable networks suitable for Noisy Intermediate-Scale Quantum (NISQ) devices, characterised by a limited number of qubits and susceptibility to noise 
\cite{TacchinoPerceptron2019, preskill2018nisq}.
The crucial step of encoding classical data into quantum states has been rigorously analysed; \textit{concurrently}, Schuld-Killoran \cite{schuld2019hilbert} and Havlíček \emph{et~al.}\ \cite{HavlicekKernel2019} independently formalised these encodings as \emph{quantum feature maps} that underpin kernel methods and variational classifiers. Cong, Choi, and Lukin introduced a Quantum Convolutional Neural Network (QCNN)  that uses only $\mathcal{O}(\log N)$ variational parameters for $N$ qubits and can recognise quantum phases \cite{CongChoiLukin}.

Variational quantum algorithms emerged as a natural progression in this field. Farhi and Neven proposed a quantum neural network (QNN) suitable for near-term processors \cite{FarhiNeven}. Mitarai \emph{et~al.} introduced quantum circuit learning, demonstrating how parameterised quantum circuits can approximate nonlinear functions \cite{MitaraiVQC2018}. Pérez-Salinas \emph{et~al.}\ formalised \emph{data re-uploading}—interleaving repeated input encoding with trainable blocks—and proved this strategy turns a shallow circuit into a universal classifier \cite{PerezReupload2020}. Addressing practical concerns, McClean \emph{et~al.} highlighted barren‐plateau trainability issues \cite{McCleanBarren2018}, and subsequent work demonstrated gradient-preserving circuit designs \cite{CerezoBarren2021} and noise-resilient optimisation strategies \cite{GentiniNoise2020}.

Recent efforts in QML have focused on integrating quantum components into classical neural network architectures.
Liu \emph{et~al.} embedded PQCs into classical convolutional blocks, yielding the hybrid quantum-classical convolutional neural network (QCCNN) and noting its robustness to moderate hardware noise \cite{LiuQCCNN2021,NoiseTolerantPQC}. Henderson \emph{et~al.} developed the Quantum Convolutional (Quanvolutional) Neural Network (QNN) by incorporating quantum layers with arbitrary quantum filters into CNN pipelines \cite{henderson2020quanvolution}, while Mari’s PennyLane demo distilled this idea to
a single quantum layer whose outputs feed a shallow artificial neural network \cite{mari2020quanvolution}. Beyond classification, QML has also been explored for generative modelling \cite{LloydWeedbrook2018}  and the community has begun to formalise benchmarking practice, e.g.\ Bowles
\emph{et~al.}\ on the pitfalls and design principles of QML benchmarks
\cite{BowlesBenchmark2024}.

Building upon the work of Henderson \emph{et~al.} \cite{henderson2020quanvolution} and Mari \cite{mari2020quanvolution}, Riaz \emph{et~al.} introduced the concept of \emph{quantum pre-processing filters} (QPFs), demonstrating their potential to improve image classification accuracy within a simple quantum kernel filter combined with a small neural network model \cite{riaz2023development}. However, despite the demonstrated utility of QPFs, the underlying reasons for their variable performance across different datasets, sometimes leading to improved accuracy and other times to degradation, remain largely unexplored. Understanding these influencing factors is critical for the robust development and deployment of hybrid quantum-classical architectures.

To address this knowledge gap, we investigated the factors contributing to the performance of QPFs, with a specific focus on their impact on validation accuracy. Our primary hypothesis centred on the role of entanglement. To systematically examine this, we introduced the concept of \emph{spatial symmetry} within the QPF circuit, which governs the qubit entanglement patterns introduced by combining rotation gates with CNOT gates. We took the exact QPF circuit designed by Riaz \emph{et~al.} and systematically modified the control-target qubit configurations of the CNOT gates. Through experimentation with 24 combinations, we identified three distinct spatial symmetries: \emph{diagonal}, \emph{vertical}, and \emph{horizontal} (see Section \ref{sec:Method} for detailed circuit configurations). Furthermore, we explored the relationship between classification accuracy and the degree of entanglement generated by each QPF symmetry for various datasets through the lens of \emph{Von Neumann entropy} \cite{BengtssonZyczkowski2017}. Our approach utilises a hybrid quantum-classical framework suitable for NISQ devices.
 
This paper presents our findings on the impact of QPF spatial symmetries on image classification performance and their correlation with entanglement levels. We detail the experimental setup and methodology in Section \ref{sec:Method}, present and discuss the classification results and entanglement analysis in Section \ref{sec:Results and Discussion}. Finally, Section \ref{sec:conclusion} concludes the paper with a summary of key insights and directions for future research.




\section{Method}\label{sec:Method}

\subsection{Hybrid Quantum–Classical Architecture and Symmetry Variants}\label{subsec:Hybrid Quantum–Classical Architecture and Symmetry Variants}

In this section, we describe the proposed hybrid quantum-classical architecture for image classification that we used to reveal intrinsic quantum hidden features from images. It utilises a quantum preprocessing filter and its variations, where permutations of circuit components are explored for quantum feature extraction, followed by a classical neural network for classification.

\begin{figure}[ht]
\centering
\includegraphics[width=\linewidth]{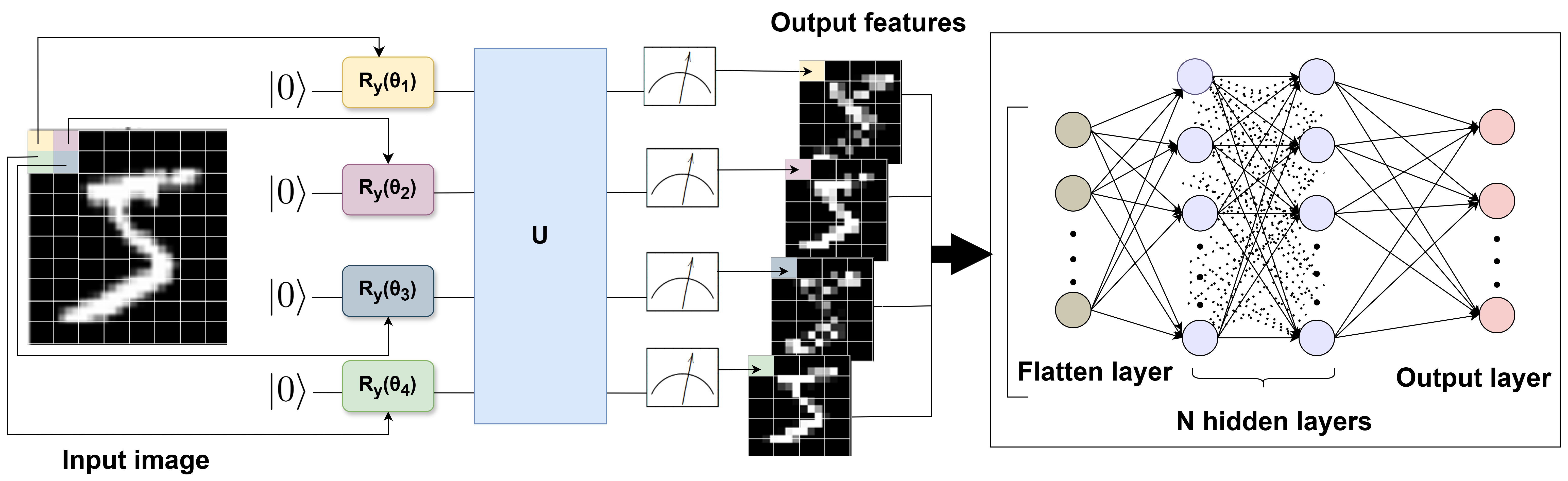}
\caption{A schematic demonstration of how the quantum circuit is used to extract the quantum information hidden in the images. We use a \(2 \times 2\) kernel that strides by 2 pixels without overlapping. The outcome of each channel is then used as the features to train a classical neural network.}
\label{fig:cnn}
\end{figure}

The filter is designed as a $2 \times 2$ quantum kernel, convolved with the image by applying it to non-overlapping $2 \times 2$ patches. This enables per-patch quantum feature extraction across the whole image, see Fig. \ref{fig:cnn}.

To evaluate this architecture, we use open-source grayscale datasets that are freely available. Each dataset provides grayscale images as $N \times N$ matrices. The images were directly processed for quantum feature extraction.

\begin{figure}[h]
\centering
\includegraphics[width=\linewidth]{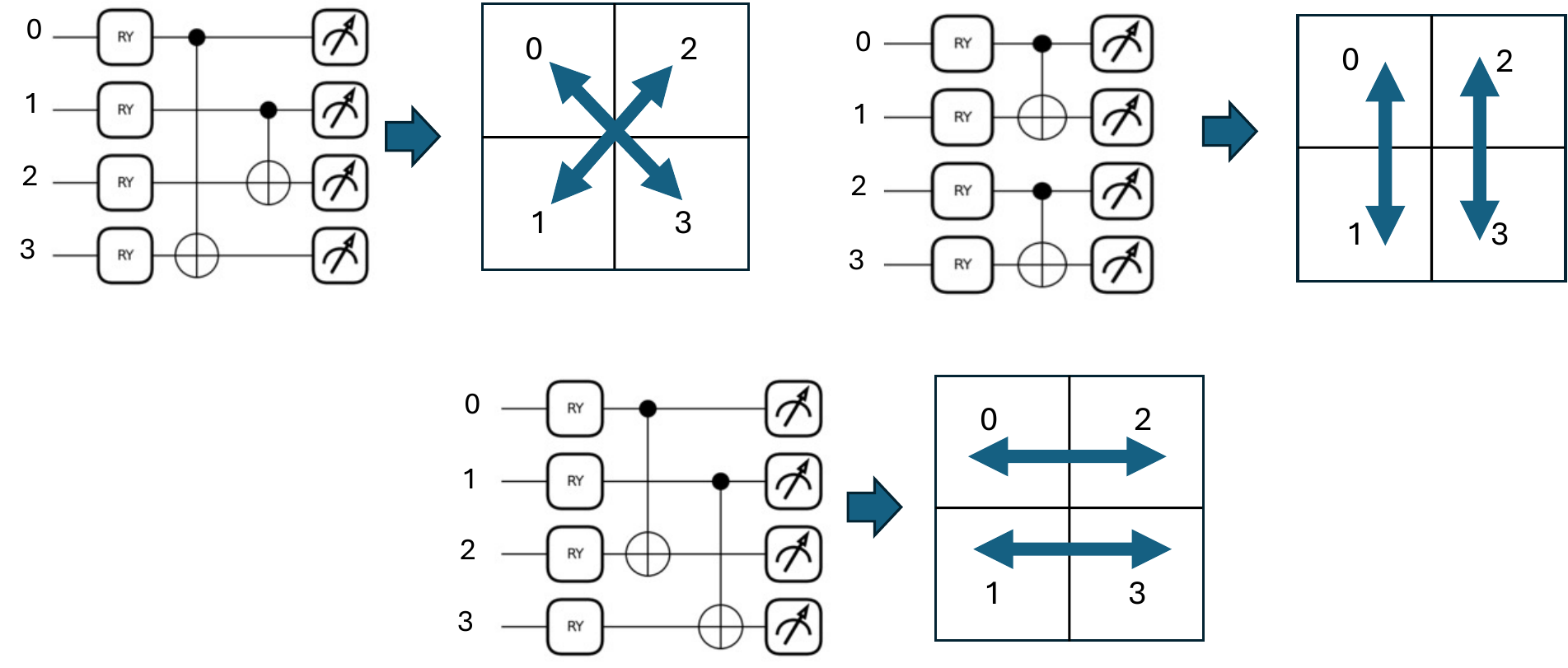}
\caption{Illustration of Possible entangling patterns for a \(2\times2\) quantum kernel.
\textbf{Upper left:} \emph{diagonal} symmetry, where qubits \((0,3)\) and \((1,2)\) are entangled;
\textbf{upper right:} \emph{vertical} symmetry, entangling qubits \((0,1)\) and \((2,3)\);
\textbf{bottom:} \emph{horizontal} symmetry, entangling qubits \((0,2)\) and \((1,3)\).
Each CNOT pair acts on an isolated two-qubit subset, so the same 2-qubit circuit can be executed in parallel across many patches or run on hardware that possesses only two qubits, making the scheme practical for NISQ devices.}
\label{fig:symmetry}
\end{figure}

The objective of this study is to determine whether classical neural networks can learn from features generated by quantum entanglement and to investigate how kernel design affects model accuracy. To test this, we enhanced a prior quantum circuit design \cite{riaz2023development} with structural changes to explore entanglement effects. Our version utilises two CNOT gates to establish qubit correlation, potentially capturing relationships between pixels that classical models may overlook. We introduce {spatial symmetry} to describe rearrangements of CNOT gates.
We observed that such an implementation using two CNOT gates has three possible symmetries for spatial pixel entanglement: diagonal, vertical, and horizontal. Figure \ref{fig:symmetry} demonstrates how the circuit can be arranged to extract the entanglement information from the image's pixels convolved with the quantum kernel. 
As shown in the circuit of Figure \ref{fig:symmetry}, each pair of qubits is entangled independently, so the kernel can run on hardware with only two qubits, and each \(2 \times 2\) image patch can be processed in parallel on multiple, identical qubit pairs—an attractive feature for NISQ-era devices where both qubit count and coherence time are limited.



\vspace{5mm}

The double-headed arrows in Fig.~\ref{fig:symmetry} indicate that every entangling pattern can be executed in both control–target directions.  
Swapping these roles lets us test whether the CNOT gate itself introduces any asymmetry.  
We confirm that it does not: a circuit that uses \(\mathrm{CNOT}(0,3)\) and \(\mathrm{CNOT}(1,2)\)—linking pixels 0–3 and 1–2—achieves the \emph{same} validation accuracy as the circuit with the directions reversed, \(\mathrm{CNOT}(3,0)\) and \(\mathrm{CNOT}(2,1)\).

Each $N \times N$ image is processed by sliding the $2 \times 2$ quantum kernel. The images are normalised to pixel values to be in the range of $[0,\pi]$.
For each patch, four pixel values are encoded into four qubits using angle embedding via $R_y(\theta)$ gates, where $\theta$ is the normalised pixel value. Measurements from each qubit are collected, producing four quantum feature maps with dimensions $(N/2) \times (N/2)$. 

\begin{figure}[ht]
\centering
\includegraphics[width=\linewidth]{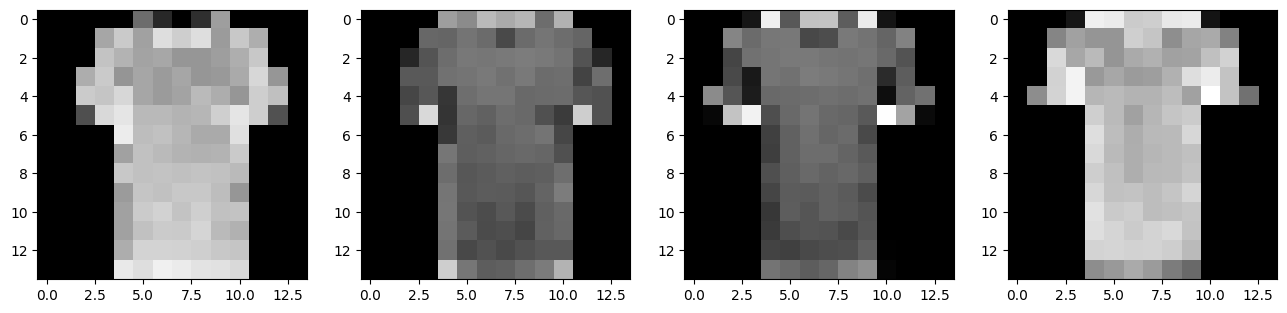}
\caption{An example of the quantum filter outcome for the Fashion MNIST dataset}
\label{fig:qchannels}
\end{figure}

Following encoding, pairwise CNOT gates introduce correlations among qubits. 
This simple design aims to test the hypothesis that specific CNOT configurations can potentially enhance feature separability through entanglement.

We experimented with different CNOT pairings, e.g., CNOT(0,1) with CNOT(2,3), and mirrored CNOT(1,0) with CNOT(3,2), as well as 24 unique pixel-to-qubit permutation mappings. 
All quantum circuits were simulated with the open-source \textit{PennyLane} software library (v 0.35.1) \cite{PennyLane}.

The resulting quantum feature channels (see Fig. \ref{fig:qchannels}) were fed into a fully connected neural network built in TensorFlow for classification. We trained for 60 epochs with a batch size of 124; the model was compiled with the Adam optimiser, sparse-categorical cross-entropy loss, and accuracy as the evaluation metric. Training employed the \texttt{ReduceLROnPlateau} callback with a patience of 15 epochs (learning rate reduced by a factor of 0.1 whenever validation accuracy failed to improve).

We also explore the performance of the filter as the depth of the neural network increases. We conducted our experiments with architectures comprising zero, one, two, and three hidden layers. For detailed results, refer to Section \ref{sec:Results and Discussion}. See Table \ref{tab:nn_architectures} for detailed information on the NN architecture. 


\begin{table}[h]
\centering
\caption{Classical Neural Network Models with Incremental Depth in Hidden Layers}
\label{tab:nn_architectures}
\begin{tabularx}{\textwidth}{c >{\raggedright\arraybackslash}X}
\toprule
\textbf{Number of Hidden Layers} & \textbf{Model Architecture} \\
\midrule

\multirow{2}{*}{0} & Input Layer (Flatten) \\
                  & Classification Layer [Dense(None, 10), softmax] \\
\midrule

\multirow{3}{*}{1} & Input Layer (Flatten) \\
                  & Hidden Layer 1 [Dense(None, 512), relu] \\
                  & Classification Layer [Dense(None, 10), softmax] \\
\midrule

\multirow{4}{*}{2} & Input Layer (Flatten) \\
                  & Hidden Layer 1 [Dense(None, 512), relu] \\
                  & Hidden Layer 2 [Dense(None, 256), relu] \\
                  & Classification Layer [Dense(None, 10), softmax] \\
\midrule

\multirow{5}{*}{3} & Input Layer (Flatten) \\
                  & Hidden Layer 1 [Dense(None, 512), relu] \\
                  & Hidden Layer 2 [Dense(None, 256), relu] \\
                  & Hidden Layer 3 [Dense(None, 128), relu] \\
                  & Classification Layer [Dense(None, 10), softmax] \\
\bottomrule
\end{tabularx}
\end{table}

All source code used to generate the results in this paper is publicly available at\\[2pt]
\url{https://github.com/TheOpenSI/QML-QPF}.

\subsection{Entanglement Quantification}\label{subsec:Entanglement Quantification}
    
    We hypothesised that the degree of entanglement the QPF, entanglement between pixel values in the images for each dataset that can be filtered by the QPF and used as features to train the NN, generates might be what makes some of the QPF-enhanced models do better—or worse—than the plain NN: there might be a relationship between the performance and entanglement.
      
    To test this hypothesis, we calculated the state vector of the two-qubit system used in the filter as a function of the input parameter from the embedding layer.
    After the two single–qubit $R_y(\theta_i)\otimes R_y(\theta_j)$ rotations and the CNOT\,$(q_i,q_j)$, the algebraic form of the two-qubit state reads
\[
|\psi(\theta_i,\theta_j)\rangle
 =A\,|00\rangle + B\,|01\rangle + C\,|10\rangle + D\,|11\rangle,
\]
where the four real amplitudes are
\[
\begin{aligned}
A&=\cos\frac{\theta_i}{2}\cos\frac{\theta_j}{2},
&\quad
B&=\cos\frac{\theta_i}{2}\sin\frac{\theta_j}{2},\\
C&=\sin\frac{\theta_i}{2}\sin\frac{\theta_j}{2},
&\quad
D&=\sin\frac{\theta_i}{2}\cos\frac{\theta_j}{2},
\end{aligned}
\]
and the angles are set by the (normalised) pixel values, $\theta_k = \pi\,x_k\in[0,\pi]$.
Because the global state is pure, tracing out either qubit gives a
$2\times 2$ reduced density matrix that can then be used to define the level of entanglement of our quantum system by using the  \textit{Von Neumann entropy}
\[
S(\rho_A) \;=\; -\operatorname{Tr}\!\bigl[\rho_A\log_2\rho_A\bigr],
\]
with $\rho_A = \operatorname{Tr}_{B} |\psi\rangle\!\langle\psi|$.

For two qubits, $S$ ranges from $0$ (separable) to $1$ (maximally
entangled); it coincides with the \emph{entanglement of formation} ($E_f(|\psi\rangle_{AB}) = S(\rho_A) = S(\rho_B)$) in this dimension, so no extra conversion factors are needed.
\vspace{2mm}

\begin{figure}[H]
    \centering
    \includegraphics[width=0.8\linewidth]{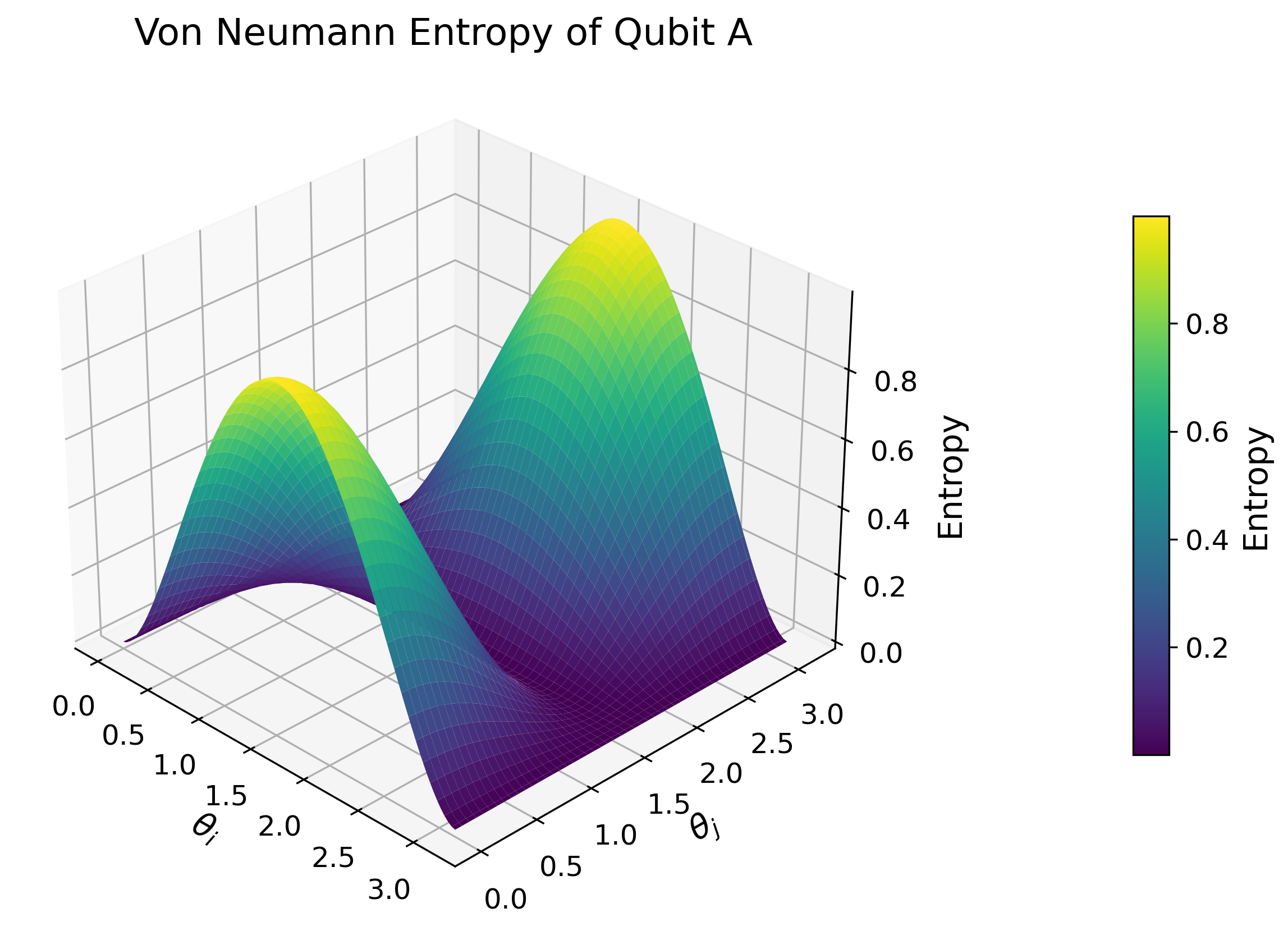} 
    \caption{Von Neumann Entropy of Qubit A in function of \(\theta_i \) and \(\theta_j \) }
    \label{fig:von_neumann_entropy} 
\end{figure}

To demonstrate the full levels of entanglement the QPF can extract, we scanned \(\theta_i,\theta_j \in [0,\pi]\) in \(0.01\pi\) steps and plotted the single-qubit von Neumann entropy \(S(\rho_A)\) over the resulting grid (Fig.~\ref{fig:von_neumann_entropy}).  
The 3-D surface shows two pronounced elongated peaks: they occur when the control qubit is near a balanced superposition \((\theta_i \approx \frac{\pi}{2})\) while the target angle is displaced from \(\frac{\pi}{2}\).  
Specifically, the points \((\theta_i,\theta_j) = (\frac{\pi}{2},0)\) \emph{and} \((\frac{\pi}{2},\pi)\) both lie on these peaks and correspond to maximally entangled Bell states, $\frac{1}{\sqrt{2}}(|00\rangle  + |11\rangle )$; $\frac{1}{\sqrt{2}}(|01\rangle  + |10\rangle )$, \cite{NielsenChuang2010} that can be reached within QPF designs.

\FloatBarrier
\section{Results and Discussion}\label{sec:Results and Discussion}


We have tested the 24 QPF permutations in a few different datasets \textit{eMNIST Digits} \cite{cohen2017emnist}, \textit{Fashion MNIST} \cite{xiao2017fashion}, \textit{MNIST} \cite{LeCunMNIST1998}, \textit{PneumoniaMNIST}, \textit{OCTMNIST}, and \textit{BreastMNIST} \cite{Yang2023MedMNIST}.


We present here box plots of validation accuracy across two datasets—\textit{eMNIST Digits}, and \textit{Fashion MNIST}—for neural network models incorporating QPF using three symmetry types: diagonal, vertical, and horizontal. For each symmetry, models were evaluated with four different architectures, varying the number of hidden layers from 0 to 3, resulting in 8 configurations per symmetry and 24 total configurations per dataset. Each box plot represents the validation accuracy of the last 15 epochs for a given configuration, joining across all configurations of that symmetry. Models without QPF were included as baselines for comparison. 

\begin{figure}[H]
    
    \includegraphics[width=\textwidth]{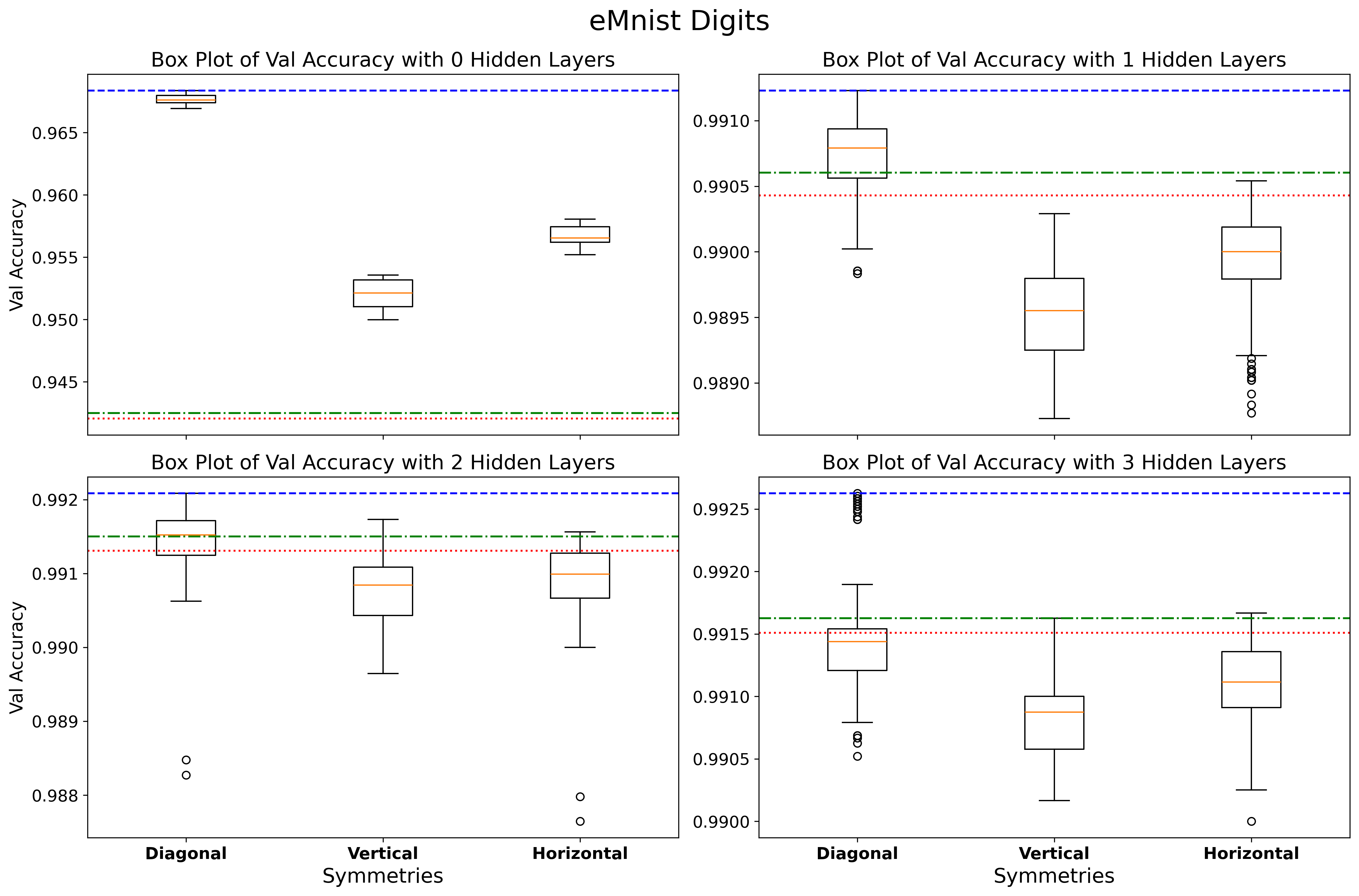}
    \caption{Validation accuracy boxplots for the eMNIST Digits dataset across QPF symmetries and layer depths. The blue dashed line indicates the maximum validation accuracy achieved across all QPF-based models. The green dot-dashed and red dotted lines represent the maximum and mean validation accuracies, respectively, of the plain neural network model.}

    \label{fig:emnist_digits_boxplot}
\end{figure}


Figure~\ref{fig:emnist_digits_boxplot} shows our results for the eMNIST Digits dataset. With no hidden layers, every symmetry-enhanced model surpasses the plain baseline; the diagonal filter yields the highest validation accuracy, followed by the horizontal and then the vertical symmetries. Adding a single hidden layer improves all curves, and the baseline briefly surpasses the horizontal and vertical QPF symmetries, although it still lags behind the diagonal symmetry. A second hidden layer raises all accuracies again, where the diagonal symmetry remains ahead. In contrast, the horizontal and vertical accuracies converge, and the baseline edges pass both but cannot match the diagonal. Introducing a third hidden layer yields only marginal gains; the diagonal symmetry remains dominant, and the baseline plateaus above the horizontal and vertical axes.\\

The sudden jumps observed in the box plots stem from the \texttt{ReduceLROnPlateau}—learning rate strategy. When validation accuracy stagnates for 15 epochs, the learning rate is reduced by 10\%, potentially resulting in a surge in performance. The clustering or overlap of some outliers further supports the interpretation that learning was temporarily stalled before resuming.

\begin{figure}[H]
    \centering
    \includegraphics[width=\textwidth]{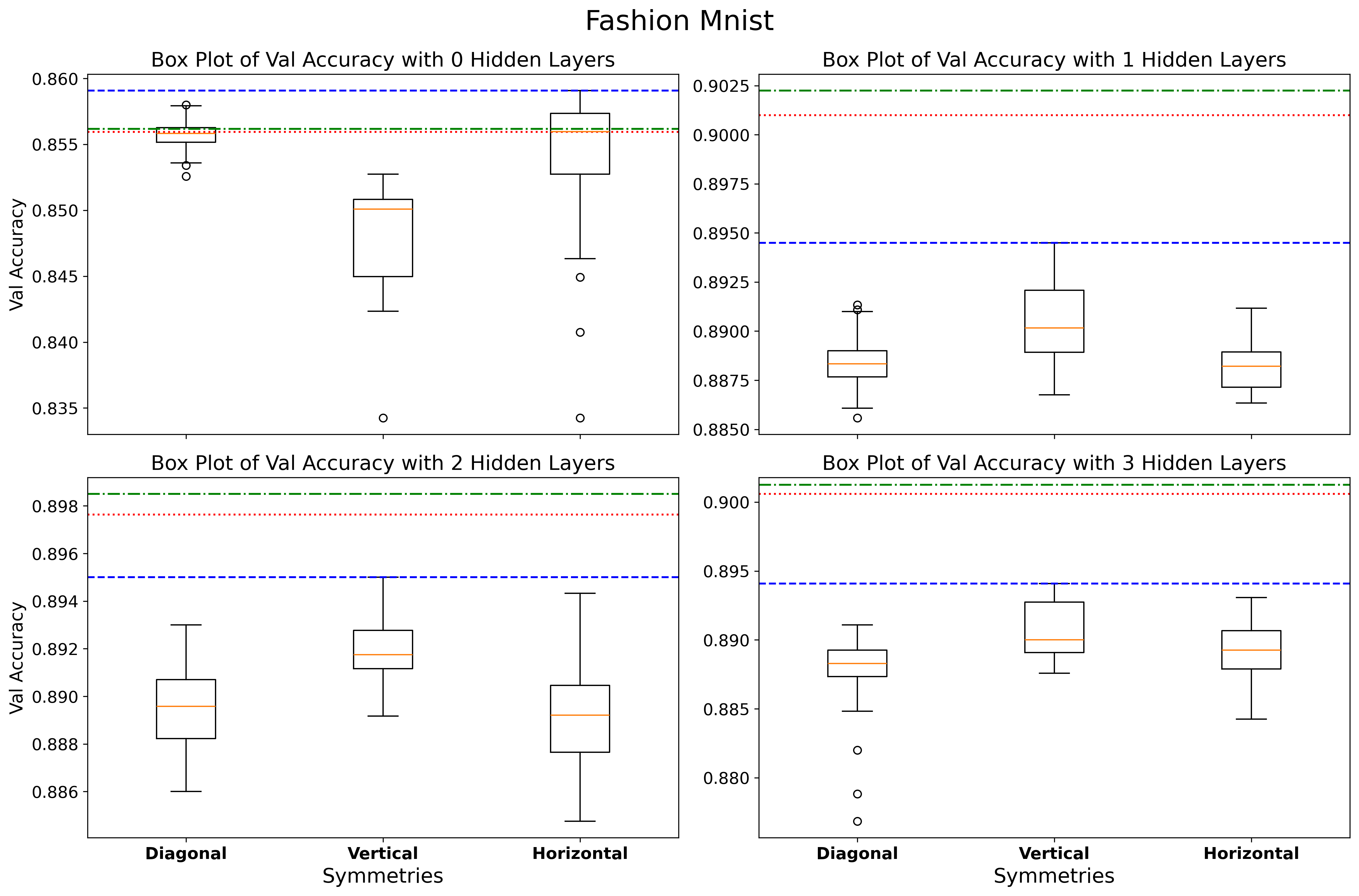}
    \caption{Validation accuracy boxplots for the Fashion MNIST dataset across QPF symmetries and layer depths. The dashed line indicates the maximum validation accuracy achieved across all QPF-based models. The dot-dashed and dotted lines represent the maximum and mean validation accuracies, respectively, of the plain neural network model.}
    \label{fig:fashion_mnist_boxplot}
\end{figure}

The pattern in Figure~\ref{fig:fashion_mnist_boxplot} when using  Fashion MNIST is noticeably different. With zero hidden layers, the horizontal symmetry performs best among all models, the diagonal symmetry offers a slight improvement over the baseline, while the vertical symmetry underperforms both the other QPF variants and the plain network. Once a hidden layer is introduced, overall accuracy climbs, but the plain network now overtakes every QPF symmetry; within the QPF symmetries, the vertical symmetry is the one which has the most impact, a reversal from the 0-hidden layer scenario, which can be seen as an indication that when combining the QPF with 0-hidden layer the NN is not able to learn the difference between the QPF symmetries for this specific dataset. 
A second hidden layer keeps the vertical filter narrowly ahead of the other symmetries, yet the baseline—despite a small dip—remains the overall leader. Depth beyond two layers yields no substantive gains: performance stabilises, the vertical symmetry continues to top the QPF symmetries, and the baseline model finishes with the highest accuracy.

\begin{figure}[H]
  \centering
  \includegraphics[width=\textwidth]{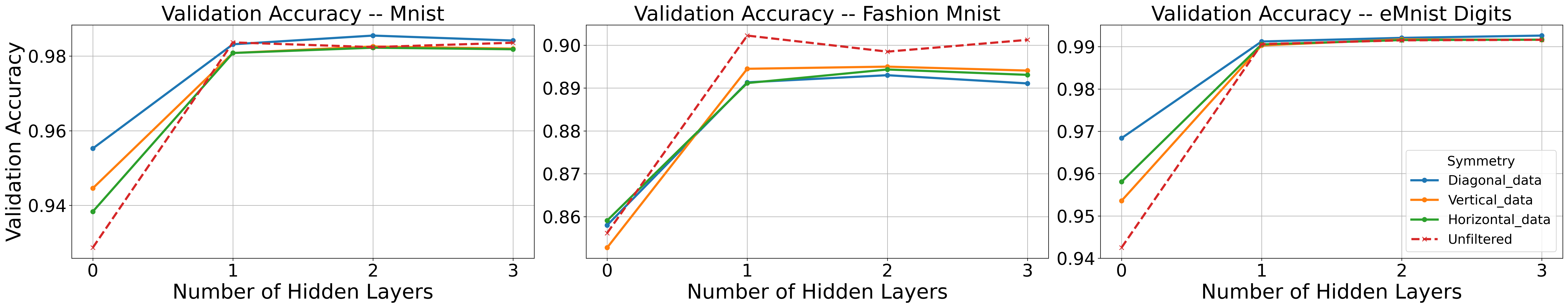}
  \caption{Validation‐accuracy trends for MNIST (left), Fashion MNIST (middle), and eMNIST Digits (right) across 0–3 hidden layers. Each plot compares diagonal, vertical, and horizontal QPF symmetries with an unfiltered baseline.}
  \label{fig:val‐trends}
\end{figure}

\vspace{2mm}
To complement the box‐plot analysis, we present line plots (Fig.~\ref{fig:val‐trends}) that track validation accuracy(the max of all symmetries) as a function of network depth (0–3 hidden layers) for the \textit{MNIST}, \textit{Fashion~MNIST}, and \textit{eMNIST Digits} datasets. This exemplifies the pattern observed in all datasets we have tested.  
Each curve compares a baseline model without QPFs
to models that apply QPFs with three symmetry types—diagonal, vertical, and horizontal.\\

\vspace{-5mm}

Table \ref{tab:layers_qpf_nn} distils the box-plot and trend-line results into a single snapshot of peak validation accuracy for every dataset–depth pair. Three clear patterns emerge. First, diagonal symmetry dominates the handwriting datasets: it edges out both the baseline and the other two symmetries on MNIST and eMNIST Digits, regardless of depth. Second, Fashion MNIST breaks that rule—vertical symmetry gives the highest QPF accuracy once at least one hidden layer is present, though the plain network still finishes on top. Third, for the three medical-imaging benchmarks (PneumoniaMNIST, OCTMNIST, BreastMNIST), the horizontal kernel is consistently best, and its relative gain grows with network depth. These results reinforce the message that each dataset has a “preferred’’ spatial symmetry.

\begin{table}[h]
\centering
\caption{Validation accuracy in \% by Dataset and Number of Hidden Layers. The best performing symmetries are: Diagonal : MNIST, eMNIST Digits; Vertical: Fashion MNIST ; Horizontal: PneumoniaMNIST,
OCTMNIST, BreastMNIST}
\label{tab:layers_qpf_nn}
\begin{tabularx}{\textwidth}{l *{8}{>{\centering\arraybackslash}X}}
\toprule
\textbf{Dataset} & \multicolumn{8}{c}{\textbf{Number of Hidden Layers}} \\
\cmidrule(lr){2-9}
& \multicolumn{2}{c}{0} & \multicolumn{2}{c}{1} & \multicolumn{2}{c}{2} & \multicolumn{2}{c}{3} \\
\cmidrule(lr){2-3} \cmidrule(lr){4-5} \cmidrule(lr){6-7} \cmidrule(lr){8-9}
& QPF & NN & QPF & NN & QPF & NN & QPF & NN \\
\midrule
MNIST          & 95.53 & 92.87 & 98.32 & 98.37 & \textbf{98.55} & 98.24 & 98.42 & 98.36 \\
Fashion~MNIST  & 85.91 & 85.62 & 89.45 & \textbf{90.22} & 89.50 & 89.85 & 89.41 & 90.13 \\
eMNIST~Digits  & 96.84 & 94.25 & 99.12 & 99.06 & 99.21 & 99.15 & \textbf{99.26} & 99.16 \\
PneumoniaMNIST & 97.03 & 96.28 & 97.35 & 97.03 & 97.56 & \textbf{97.77} & 97.56 & 96.82 \\
OCTMNIST       & 69.04 & 65.77 & 83.19 & 81.70 & 84.51 & 82.95 & \textbf{84.79} & 83.02 \\
BreastMNIST    & 82.73 & 73.64 & 86.36 & 80.00 & 84.55 & 82.73 & \textbf{88.18} & 80.91 \\
\bottomrule
\end{tabularx}
\end{table}


Overall, we observed that most of the datasets used in this study benefit from improved accuracy when leveraging one of the symmetries of the QPF, with the notable exception of Fashion MNIST. We explore a possible explanation for this anomaly in the following sections.



\subsection{Datasets' entanglement level }

To look for correlation between the performance of the NN models that use the QPFs and the level of entanglement generated by these QPFs, we recorded the entropy of the QPF-encoded states for all images in each dataset, then averaged over the entire dataset. We repeated this for every dataset presented in this study. The resulting value for the best-performing configuration for each symmetry is shown in Table~\ref{tab:avg-entanglement}.


\begin{table}[h]
\centering
\caption{Average single-qubit von Neumann entropy (bits) for the best-performing configuration of each symmetry.
Numbers in parentheses are standard deviations across images. The numbers in bold represent the model with the highest validation accuracy in \% during training.}
\label{tab:avg-entanglement}
\begin{tabularx}{\textwidth}{c *{6}{>{\centering\arraybackslash}X}}
\toprule
\multirow{2}{*}{\textbf{Dataset}} & \multicolumn{2}{c}{\textbf{Diagonal}} & \multicolumn{2}{c}{\textbf{Vertical}} & \multicolumn{2}{c}{\textbf{Horizontal}} \\
& Entropy (std) & Accuracy (\%) & Entropy (std) & Accuracy (\%) & Entropy (std) & Accuracy (\%) \\
\midrule
MNIST          & \textbf{0.106\,(0.032)} & \textbf{98.4}  & 0.107\,(0.032) & 98.1 & 0.098\,(0.028) & 98.1 \\
Fashion~MNIST  & 0.286\,(0.102) & 89.1  & \textbf{0.286\,(0.102)} & \textbf{89.4} & 0.249\,(0.090) & 89.3 \\ 
eMNIST~Digits  & \textbf{0.154\,(0.043)} & \textbf{99.2}  & 0.154\,(0.043) & 99.1 & 0.147\,(0.045) & 99.1 \\ 
PneumoniaMNIST & 0.444\,(0.064) & 97.3  & 0.444\,(0.064) & 97.4 & \textbf{0.438\,(0.062)} & \textbf{97.5} \\ 
OCTMNIST       & 0.370\,(0.068) & 83.9  & 0.361\,(0.070) & 83.7 & \textbf{0.370\,(0.068)} & \textbf{84.7} \\ 
BreastMNIST    & 0.422\,(0.070) & 87.2  & 0.422\,(0.070) & 86.3 & \textbf{0.387\,(0.069)} & \textbf{88.1} \\ 
\bottomrule
\end{tabularx}
\end{table}

Within–dataset analysis shows no clear link between average entanglement and validation accuracy; the three symmetries differ by at most ${\sim}0.01$--$0.04$\ bits. It is impossible to identify a difference between symmetries in entropy; nevertheless, their kernels produce different validation accuracies, as shown in bold, representing the symmetry with the best validation accuracy during training.   

We also observe that the relationship between the level of entanglement and the filter's performance across different datasets is not well-defined. 
In some scenarios using the QPF, a slightly higher entropy yields improved classification, while in others, a slightly lower entropy results in improved classification compared to a plain NN (see Table \ref{tab:layers_qpf_nn}). Therefore, we can say that there's no guaranteed outcome based solely on the level of entanglement.  

Future research should investigate whether an optimal “sweet spot” for entanglement may exists that balances performance and how this interacts with factors such as dataset complexity and network capacity. While developing a state-of-the-art quantum neural network model is beyond the scope of this work, gaining a deeper understanding of these quantum filters could pave the way for hybrid models that integrate quantum and classical filtering techniques.

\subsection{General Observations} 
Across all datasets, increasing model depth consistently improves validation accuracy. However, the impact of QPFs symmetries depends on the dataset. The diagonal symmetry QPF consistently performs well, particularly on MNIST and eMNIST Digits, suggesting that it effectively captures key features in simpler datasets.
The medical imaging datasets (PneumoniaMNIST, OCTMNIST, BreastMNIST) show a clear preference for horizontal symmetry, with this QPF variant consistently achieving the best performance across all three medical datasets, and its relative advantage growing with network depth.
In contrast, for the more complex Fashion MNIST dataset, the baseline neural network outperforms all QPF-enhanced models.

Nevertheless, for the Fashion MNIST dataset, the vertically symmetric QPF still contributes valuable representational power: once the network has at least one hidden layer, the vertical kernel persistently outperforms the diagonal and horizontal kernels and remains the closest QPF competitor to the baseline. This pattern suggests that, even though vertical symmetry does not surpass the plain model in headline accuracy, it does succeed in isolating important visual features specific to Fashion~MNIST--that other symmetries fail to capture as effectively. The result highlights two complementary points: (i) raw accuracy alone may mask the qualitative benefits of a symmetry that distils informative features, and (ii) selecting or designing a QPF kernel must be guided by both the statistical structure of the data and the capacity of the classifier to leverage the extracted features.\\
 In particular, the presence of outliers in the boxplots is not indicative of anomalies but rather a consequence of the \texttt{ReduceLROnPlateau} learning rate scheduler, which enables the model to escape stagnation and resume effective learning.\\

These results lead to several important insights. First, the effectiveness of a given symmetry is highly dataset-dependent. Diagonal symmetry yields strong performance in MNIST and eMNIST Digits, while vertical symmetry appears better suited for Fashion MNIST once the model is sufficiently deep, and horizontal symmetry demonstrates superior performance for PneumoniaMNIST, OCTMNIST, and BreastMNIST. This suggests that no single QPF symmetry generalises universally across datasets, and that the optimal symmetry choice correlates with the underlying patterns most relevant to each dataset.

Second, model depth plays a critical role. Shallow networks are less capable of leveraging the symmetries in filtered data, particularly in complex datasets. However, as depth increases, the network's ability to extract and amplify the underlying structure improves. In Fashion~MNIST, vertical symmetry never quite surpasses the plain model in raw accuracy, yet it reliably isolates intricate features that diagonal and horizontal filters overlook. This outcome highlights that headline accuracy can obscure the qualitative benefit of a symmetry that distils information, and emphasises that kernel choice must balance the data’s statistical structure with the classifier’s capacity to exploit the extracted features.
    

These findings indicate that future work should develop a framework—or selection strategy—that automatically matches each dataset’s characteristics to the most effective QPF symmetry (or kernel). Such a tool would significantly improve the usability and performance of hybrid quantum-classical architectures in real-world applications. 
    






%



\section{Conclusion}\label{sec:conclusion}
This study investigated the impact of quantum pre-processing filters, with varying designs based on the spatial symmetries, on the performance of hybrid quantum-classical neural networks across different image datasets. We observed that increasing model depth consistently improved validation accuracy across all datasets. However, the efficacy of QPF symmetries proved to be highly dataset-dependent. Diagonal symmetry QPFs consistently performed well on simpler datasets like MNIST and eMNIST Digits, suggesting their effectiveness in capturing key features. For medical imaging datasets (PneumoniaMNIST, OCTMNIST, BreastMNIST), horizontal symmetry consistently yields the best performance.  
In contrast, for the more complex Fashion MNIST dataset, the classical baseline neural network generally outperformed QPF-enhanced models, though the vertical symmetry QPF(for hidden layers $\ge 1$) demonstrated valuable representational power, isolating specific visual features that other symmetries failed to capture.\\

Our findings highlight that raw accuracy alone may not fully reflect the qualitative benefits of a symmetry that effectively distils informative features. Furthermore, the choice and design of a QPF kernel must consider both the statistical structure of the data and the classifier's capacity to leverage the extracted features. While entanglement levels were quantified, no strong correlation was found between model accuracy and the degree of entanglement in the filtered images, suggesting that performance gains cannot be solely attributed to this factor. These results underscore the need for future work to develop frameworks for automatically matching dataset characteristics to the most effective QPF symmetry or kernel, thereby enhancing the usability and performance of hybrid quantum-classical architectures in real-world applications.\\

Our conclusions should be viewed in light of several limitations. First, the experiment was conducted on simulated noise-free circuits; gate errors on NISQ hardware may reduce performance. Second, the entanglement metric was restricted to pairwise Von Neumann entropy, leaving multi-qubit correlations unexplored. Addressing these issues—by benchmarking on real devices and by extending the information-theoretic analysis—will be the focus of future work.



\backmatter





\bmhead{Acknowledgements}

This work is funded under the agreement with the ACT Government, Future Jobs Fund - Open Source Institute (OpenSI) - R01553.

\bibliography{Kernel_QML_bibliography}



\end{document}